\documentclass{mem}
\usepackage{natbib}\usepackage{txfonts}\usepackage{balance}
\usepackage{graphicx}
\usepackage[a4paper]{hyperref}
\idline{75}{282}

\def\cgs{{ erg cm$^{-2}$ s$^{-1}$}}

\def\cm2{{ cm$^{-2}$}}

\begin{document}
\def\teff{$T\rm_{eff }$}
\def\kms{$\mathrm {km s}^{-1}$}

\title{Science with Simbol-X}

   \subtitle{}

\author{
F. Fiore\inst{1}
\and M.~Arnaud\inst{2}
\and U.~Briel\inst{3}
\and M.~Cappi\inst{4}
\and A.~Comastri\inst{5}
\and A.~Decourchelle\inst{2}
\and R.~Della Ceca\inst{6}
\and P.~Ferrando\inst{7}
\and C. Feruglio\inst{1}
\and R.~Gilli\inst{5}
\and P.~Giommi\inst{8}
\and A.~Goldwurm\inst{7}
\and P.~Grandi\inst{4}
\and P.~Laurent\inst{7}
\and F.~Lebrun\inst{7}
\and G.~Malaguti\inst{4}
\and S.~Mereghetti\inst{9}
\and G.~Micela\inst{10}
\and G.~Pareschi\inst{6}
\and E.~Piconcelli\inst{1}
\and S.~Puccetti\inst{8}
\and J.P.~Roques\inst{11}
\and G.~Tagliaferri\inst{6}
\and C.~Vignali\inst{12}
          }

  \offprints{F. Fiore}

\institute{
INAF-OARoma, Via Frascati 33, I-00040 Monteporzio, Italy
\and
Laboratoire AIM, DAPNIA/Service d'Astrophysique - CEA/DSM - CNRS
 - Universit\'{e} Paris Diderot, B\^{a}t. 709, CEA-Saclay, F-91191
 Gif-sur- Yvette Cedex, France
\and
Max Planck Instit\"ut f\"ur Extraterrestrische Physik, 85748 Garching, Germany
\and
INAF-IASF-Bo, via Gobetti 101, 40129 Bologna, Italy 
\and
INAF-OABo, via Ranzani 1, Bologna, Italy 
\and
INAF-OABrera, via Bianchi 46, 23807 Merate, Italy
\and
UMR 7164, Laboratoire APC \& DSM/DAPNIA/Service d'Astrophysique CEA/Saclay, 
91191 Gif-sur-Yvette Cedex, France
\and
ASI ASDC c/o ESRIN, 00044 Frascati, Italy
\and
INAF-IASF-Mi, Via Bassini 15, 20133 Milano, Italy
\and
INAF-OAPa, Palazzo dei Normanni, 90134 Palermo, Italy
\and
CESR, BP 4346, 31028 Toulouse Cedex, France
\and
Dipartimento di Astronomia, Universit\'a di Bologna, via Ranzani 1, Bologna, Italy
}

\authorrunning{Fiore et al.}

\titlerunning{Science with Simbol-X}

\abstract{ Simbol-X is a French-Italian mission, with a participation
of German laboratories, for X-ray astronomy in the wide 0.5-80 keV
band. Taking advantage of emerging technology in mirror manufacturing
and spacecraft formation flying, Simbol-X will push grazing incidence
imaging up to $\sim80$ keV, providing an improvement of roughly three
orders of magnitude in sensitivity and angular resolution compared to
all instruments that have operated so far above 10 keV. This will open
a new window in X-ray astronomy, allowing breakthrough studies on
black hole physics and census and particle acceleration mechanisms. We
describe briefly the main scientific goals of the Simbol-X mission,
giving a few examples aimed at highlighting key issues of the
Simbol-X design.   } \maketitle{}

\section{Introduction}

A seminal result obtained with HEAO-1 at the end of the 70' is the
precise measure of the spectrum (from a few keV up to $\sim100$ keV) of
the isotropic, extragalactic Cosmic X-Ray Background (CXB), discovered
by Riccardo Giacconi, Bruno Rossi and collaborators during one of the
first rocket-borne X--ray experiments in 1962. The HEAO1 data showed
that the CXB energy density has a broad maximum around 30 keV, where
it is about 5 times higher than at 1 keV and 50\% higher than at 10
keV.  It was soon realized that the CXB is most likely due to the
contribution of many discrete sources at cosmological distances
\citep{sw79}. Most of these sources are active galactic nuclei,
AGN, implying that the CXB energy density provides an integral
estimate of the mass accretion rate in the Universe, and therefore of
the super-massive black hole (SMBH) growth and mass density.
Unfortunately, the integrated light from all sources detected in
HEAO1 all-sky survey could directly explain only less than 1\% of the
CXB.  Indeed, the use of collimated detectors on board first UHURU and
Ariel-V, and then HEAO1 in the 1970 decade led to the discovery of
$<1000$ X-ray sources in the whole sky.

X-ray imaging observations, performed first by {\it Einstein} and
ROSAT in the soft X-ray band below $\sim3$ keV, and then by ASCA,
BeppoSAX, XMM--Newton and Chandra up to 8-10 keV, detected tens of
thousands of X-ray sources, and resolved nearly 100\% of the CXB below
a few keV and up to 50\% at 6-8 keV. These observations increased by
orders of magnitude the discovery space for compact objects (both
Galactic neutron stars and black holes and AGN) and for thermal plasma
sources. However, they still leave open fundamental issues, such as
what is making most of the energy density of the CXB at $\sim30$ keV.

Above 10 keV the most sensitive observations have been performed so
far by collimated instruments, like the BeppoSAX PDS, and by coded
masks instruments, like INTEGRAL IBIS and Swift BAT. Only a few
hundred sources are know in the whole sky in the 10-100 keV band, a
situation recalling the pre--{\it Einstein} era at software energies.
A new window in X-ray astronomy above 10 keV must be opened, producing
an increase of the discovery space similar to that obtained with the
first X-ray imaging missions. This will be achieved by Simbol-X, a
formation flight mission currently under preparation by France and
Italy, with a participation from German laboratories. Very much like
the {\it Einstein} Observatory, this mission will have the
capabilities to investigate almost any type of X-ray source, from
Galactic and extragalactic compact sources, supernova remnant (SNR),
young stellar objects and clusters of galaxies, right in the domain
where accretion processes and acceleration mechanisms have their main
signatures.  This paper summarizes the main scientific goals of the
Simbol-X mission, putting the emphasis on the core science objectives.

\section{Main scientific objectives of the Simbol-X mission}

Taking advantage of emerging technology in mirror manufacturing
(Pareschi et al. these proceedings) and spacecraft formation flying
\cite{lamarle07}, Simbol-X will push grazing incidence imaging up to
$\sim80-100$ keV, providing an improvement of roughly three orders of
magnitude in sensitivity and angular resolution compared to all
instruments that have operated so far above 10 keV (Fig. 3 in Ferrando
et al., these proceedings, compares the predicted Simbol-X sensitivity
to that of previous experiments in the 1-100 keV band).  The very wide
discovery space that Simbol-X will uncover is particularly significant
for the advancement of two large and crucial areas in high-energy
astrophysics and cosmology:

\begin{enumerate}
\item
Black hole physics and census
\item 
Particle acceleration mechanisms. 
\end{enumerate}
 
These two broad topics define the core scientific objectives of
Simbol-X.

Because of the tight links between galaxy bulges and their central
SMBH, obtaining a complete and unbiased census of SMBH, through direct
observations at the energies where the Cosmic X-ray Background (CXB)
energy density peaks, is crucial for our understanding of the
formation and evolution of galaxies and their nuclei. Furthermore, BH
environment is the only known place in the Universe where general
relativity can be tested beyond the weak-field limit.

About particle acceleration, we are still lacking firm evidences of
hadron acceleration in astronomical sites (despite clearly seeing huge
electron accelerations), and we are still searching for the origin of
the high-energy photons and cosmic rays. Hard X-rays observations,
possibly combined with $\gamma$--ray and TeV observations, are
invaluable tools to identify the processes at work in acceleration
sites such as SNR and jets.To achieve its core scientific objectives
Simbol-X should:

\begin{enumerate}
\item[1.1] resolve at least 50\% of the CXB in the energy range where
it peaks, thus providing a more complete census of SMBH;

\item[1.2] solve the puzzle on the origin of the hard X-ray emission
from the Galactic centre, which  harbors the closest SMBH;

\item[1.3] constrain the physics and the geometry of the accretion
flow onto both SMBH and solar mass BH;

\item[1.4] map the messy environment around SMBH characterized by
the coexistence of gas components with different dynamical, physical
and geometrical properties;

\item[2.1] constrain acceleration processes in the relativistic Jets
of blazars and GRB;

\item[2.2] probe acceleration mechanisms in the strong electromagnetic and 
gravitational fields of pulsars;

\item[2.3] measure the maximum energy of electron acceleration in supernova 
remnants shocks, and search for hadron acceleration in these sites;

\item[2.4] search for and map the non-thermal emission in clusters of
galaxies, and if confirmed, determine its origin and its impact on
clusters evolution.

\end{enumerate}

In addition to these top priority objectives Simbol-X will be capable
of performing breakthrough studies on several other areas like:

\begin{enumerate}

\item
the equation of state and the magnetic field of neutron stars;
\item
nucleosynthesis in young SNR; 
\item
the formation of stars and planets;
\item
non-thermal emission of active stars;
\item
shocks in the intracluster medium pervading groups and clusters of galaxies;
\item
extended thermal plasmas in Galactic and extragalactic sources.
\end{enumerate}

\begin{figure*}[ht!]
\begin{center}
\includegraphics[height=10truecm,width=10truecm]{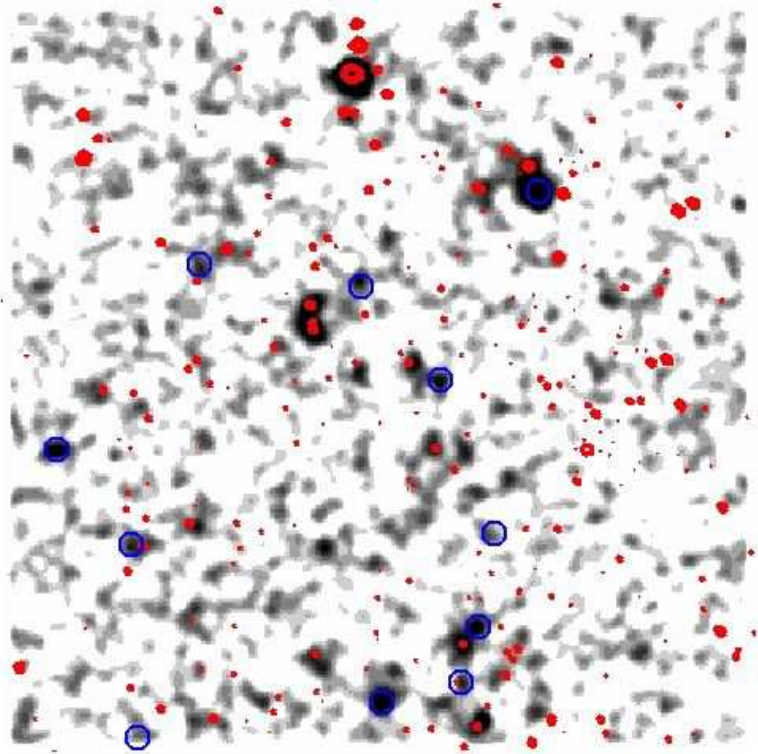}
\end{center}
\caption{\footnotesize
The combined CZT+SDD 1M sec image of the CDFS in the 10-40 keV band
(the exposure in the SDD camera is half of the total elapsed time to
account for dead time). Red contours are Chandra count rates, circles
identify the highly obscured, infrared selected sources. For
the Chandra sources we have computed the 10-40 keV fluxes
extrapolating the 2-10 keV fluxes using a spectral model consistent
with the Chandra hardness ratios. For the IR selected sources we
converted their 24$\mu$m fluxes using typical IR to X-ray unabsorbed flux
ratios, $\alpha_E=0.8$ and N$_H$ in the range $10^{24}-10^{25}$ cm$^{-2}$.  }
\label{cdfs}
\end{figure*}

Of course we cannot discuss here in detail all above topics, these are
exhaustively presented in all papers in this volume. We limit
ourselves to provide a few examples, highlighting the key issues about
the Simbol-X design with respect to possible competitors, and the
synergies with other large observational infrastructures that are
already producing data or that will produce data at the time of the
Simbol-X mission (2013-2018).

\subsection{The cosmic X-ray background and the census of SMBH black holes}

The CXB is currently regarded as the integrated output of the
accretion processes which took place during the cosmic history. These
processes led to the growth of SMBH in galactic nuclei (e.g. Marconi
et al. 2004), which we observe in an active phase in AGN and in a
quiescent phase through their dynamical effects on their surroundings,
at the centre of nearby galaxies and indeed in our own.

AGN making most of the CXB below a few keV have a spectrum much softer
than the CXB energy density spectrum, implying that the maximum at 30
keV would be missed by a factor about 3. A simple solution of this
``paradox'' was proposed again by Setti and Woltjer (1989), and
requires a population of AGN highly obscured in soft X-rays by
photoelectric absorption. The size of this population should be 2-3
times that of the unobscured AGN to reproduce the CXB spectrum (see
Comastri et al. these proceedings).  Chandra and XMM-Newton surveys
have been able to detect many obscured, Compton-thin AGN
(N$_H<10^{24}$ cm$^{-2}$). However, due to their limited band pass,
Chandra and XMM discovered only a handful of Compton-thick AGN (CT;
N$_H>10^{24}$ cm$^{-2}$).  Therefore, at present, only a few CT AGN
are known beyond the local Universe (see Della Ceca et al. these
proceedings).

According to both the most up-to-date AGN synthesis models for the CXB
\citep{gilli07}, the volume density of CT AGN should be of the same
order of magnitude of that of the unobscured and moderately obscured
AGN.  High sensitivity, hard X-ray observations like those that
Simbol-X will be able to perform in the 10-60 keV band hold the key to
uncover, and study in detail, this long sought population of CT AGN,
thus detecting directly most SMBH accretion luminosity in the
Universe. Obtaining a complete census of accreting SMBH through the
cosmic epochs is a crucial step to constrain nuclear accretion
efficiency and feedbacks on the host galaxies, which are key
ingredients toward the understanding of galaxy formation and
evolution. The Simbol-X main contributions in this field will be the
discovery and the characterization of the sources making the main
contribution to the peak of the CXB. Simbol-X will allow us to
evaluate the luminosity function of obscured AGN and its evolution,
and to measure the fraction of obscured AGN as a function of
luminosity and redshift with little observational biases. To these
purposes the following observational strategies can be envisaged:

[1.] A spectral survey of local CT Seyfert 2 galaxies and moderately
obscured QSOs previously discovered by BeppoSAX, XMM, INTEGRAL, Swift
and Suzaku. This may be accompanied by a survey of infrared bright
galaxies which have not shown strong X-ray emission below 10 keV, to
search for highly CT objects.
These observations will allow the precise measure of the absorbing
column density, and therefore the determination of the N$_H$
distribution of a large sample AGN sample, including CT objects (Della
Ceca et al. these proceedings). They will also allow us to put
constraints on the physical status of the absorbing gas and on its
covering fraction.

[2.] Deep observations to search for higher redshift CT AGN.

[3.] A serendipitous survey over a large area to search for high
luminosity CT AGN.

[4.] A survey of candidate CT AGN selected using their infrared
emission and Spitzer/Herschel surveys, see Feruglio et al. these
proceedings.


These observations will quantify the obscured AGN volume density as a
function of the Cosmic time and univocally confirm and identify the IR
selected CT AGN as hard X-ray AGN, contributing to the CXB.

\subsubsection{The CDFS: a case study}

As an example of what Simbol-X can achieve in the field extragalactic
deep survey, Fig. \ref{cdfs} shows a simulation of a 1Msec observation
of the Chandra Deep Field South area in the 10-40 keV band. We have
included in the simulation two source populations: 1) the X-ray
sources detected by Chandra below 10 keV; 2) the candidate CT AGN
selected in the mid-infrared by Fiore et al. (2007). For the former
sources we extrapolated their flux in the 10-40 keV band using a
spectral model consistent with the Chandra hardness ratios. For the IR
selected CT AGN we converted their 24$\mu$m fluxes to X-ray unobscured
fluxes using typical templates (Fiore et al. 2007). We than computed
observed 10-40 keV fluxes by assuming $\alpha_E=0.8$ and N$_H$ in the
range $10^{24}-10^{25}$ cm$^{-2}$. Note as a few IR selected AGN are
detected also by Chandra, but several others can be detected only by
Simbol-X above 10 keV. A single Simbol-X 1Msec observation will be
able to resolve about 50\% of the CXB in the 10-40 keV band. We will
be able to probe nearly all kind of AGN, from unobscured to moderately
obscured to CT AGN. Since the fraction of CT AGN rises steeply below
10$^{-14}$ \cgs (10-40 keV band) pushing the flux limit just below
this value would strongly increase the chance to discover relativaly
large samples of these elusive objects (see Comastri et al. these
proceedings for further details).

Main key issue in this field is the sharp image quality of the
Simbol-X mirror (a Point Spread Function, PSF, with half power
diameter, HPD$<$ 20 arcsec, Pareschi et al. these proceedings). Other
important factors are the relatively large field of view ($\sim 12$
arcmin diameter) and the low internal background (see Laurent et al.
these proceedings).  Already mentioned synergies are with mid and far
infrared space observatories like Spitzer, Herschel and JWST.

\subsection{AGN massive outflows: AGN feedback at work} 

In addition of obtaining a complete census of SMBH, the other key
observational ingredient to obtain a  better understanding of galaxy
formation and evolution  is the quantification of the effects of 
AGN feedbacks on their host galaxies. 

AGN can interact with the interstellar matter of their host galaxies
through at least two main channels: radiation field and
outflows. Winds and jets, both non-relativistic and relativistic, are
common in AGN, see Cappi et al. and Tavecchio et al. these
proceedings.  The two key parameters are the mass outflow rate and the
velocity structure of the outflow, because they give the energy and
the momentum involved in the flow. Blue-shifted X-ray absorption lines
detected by Chandra and XMM indicate very high velocities, up to a
fraction of the speed of light, in a few sources. A comprehensive
survey of absorption features in sizeable AGN samples in different
bins of redshift and luminosity is however still lacking. Furthermore,
the sensitivity of XMM decreases sharply above 9-10 keV, hampering the
possibility of detecting high velocity outflows in nearby AGN.
Simbol-X will be 2 to 5 times more sensitive to iron absorption
features than XMM-Newton below 10 keV, and will open the window above
10 keV. This will allow the characterization of outflows of any
velocity in statistical samples. The study of the variability of the
absorption lines will set a scale for the size and location of the
absorbing gas, and for its density, thus providing information on the
mass involved in the outflow.

Key issue in this topic is the high Simbol-X throughput between 7 an
20 keV. Other important issues are the good energy resolution provided
by the MPC detector and the broad band coverage, allowing a good
constraint on the source continuum.

\subsection{Acceleration mechanisms in supernova remnant}

Non-thermal emission was originally discovered by ASCA in the
shell-like supernova remnant (SNR) SN1006 \citep{koyama95}. We know
today that most of the young SNR show non-thermal emission at some
level. These are the best candidate sites for the acceleration of
Cosmic Rays up to $10^{15}$ eV or even higher energies. Indeed, if the
maximum energy of electron ($E_{max}$) is of this order, a strong
synchrotron emission is expected in the X-rays band, with a cut-off at
an energy $E_{cut}$ depending on $E_{max}$ and on the strength of the
magnetic field.  The magnetic field necessary for their scattering may
be generated or increased by the accelerated particles themselves (the
so called streaming instability). Evidences that this effect is at
work were claimed thanks to Chandra observations of the thickness of
the X-ray rims around the SN shock in the Tycho SNR
(Cassam-Chena{\"\i} et al. 2007).  The brightness profile of the rims
is consistent with magnetic fields a few hundred times larger than the
ISM field.  However, this interpretation of the Chandra data is not
unique. The direct measure of $E_{cut}$ will then help understanding
both which is the maximum energy of electron and the mechanism of
diffusive shock acceleration in SNR shells (see Decourchelle et
al. these proceedings).

Although we have plenty of evidences of electron acceleration in
cosmic sources, direct proofs of proton and ion acceleration are still
lacking. A direct signature of proton acceleration is GeV-TeV emission
due to pion decay. HESS and MAGIC have revealed TeV emission from a
few supernova remnants (e.g. Aharonian et al. 2006).  A large fraction
of the TeV photons are probably due to Inverse Compton (IC)
emission. To disentangle between pion decay and IC emission one needs
to estimate the expected IC emission through their synchrotron
emission observed in the X-ray range. Combining hard X-ray and
$\gamma$-ray observations holds the key to uncover pion decay
emission.

Key issue in this field is a broad energy band, extending from $\sim1$
keV to $\sim100$ keV. This will allow the separation of thermal and
non-thermal emission and the measure of $E_{cut}$. Extremely important
here is to have a large field of view, to cover large fractions of
SNR. Obvious synergies are with TeV Cherenkov telescopes and with
GLAST and Agile.

\section{Conclusions}

Thanks to the emerging technology in X-ray mirrors (e.g. multilayer
coating, see Pareschi et al. these proceedings) and spacecraft
formation flying, Simbol-X will provide a large collecting area (of
the order of 100-1000 cm$^{-2}$) from a fraction of keV up to $\sim80$
keV, thus overcoming the ``10 keV limit'' for high accuracy imaging
and spectroscopy of all past and current X-ray observatories.  This,
together to the good image quality (PSF HPD$<20$ arcsec, FWHM$<10$
arcsec), relatively large field of view (12 arcmin diameter), good
detector quantum efficiency, resolution and low internal background,
will allow breakthrough studies on black hole physics and census, and
particle acceleration mechanisms.

\begin{acknowledgements}
We acknowledge support from ASI-INAF contracts I/023/05/0, I/088/06/0 and
PRIN-MUR grant 2006-02-5203
\end{acknowledgements}

\bibliographystyle{aa}

\end{document}